\documentstyle[twocolumn,prl,aps,epsfig]{revtex}

\begin{document}
\draft
\def \beq{\begin{equation}}
\def \eeq{\end{equation}}
\def \beqarr{\begin{eqnarray}}
\def \eeqarr{\end{eqnarray}}

\twocolumn[\hsize\textwidth\columnwidth\hsize\csname @twocolumnfalse\endcsname

\title{ 
Rigorous Proof of
Pseudospin Ferromagnetism in Two-Component Bosonic Systems with 
Component-Independent Interactions
}

\author{Kun Yang$^{1,2}$ and You-Quan Li$^{2}$}

\address{
$^1$National High Magnetic Field Laboratory and Department of Physics,
Florida State University, Tallahassee, Florida 32306
}
\address{
$^2$Zhejiang Institute of Modern Physics, Zhejiang University, Hangzhou 310027,
P. R. China
}

\date{\today}

\maketitle

\begin{abstract}
For a two-component bosonic system, the components can be mapped onto a 
pseudo-spin degree of freedom with spin quantum number $S=1/2$.
We provide a rigorous proof that for a wide-range of real Hamiltonians with
component independent mass and interaction, the ground state is a ferromagnetic
state with pseudospin
fully polarized. The spin-wave excitations are studied and found to have 
quadratic dispersion relations at long wave length. 
\end{abstract}


]

There has been an explosion of interest in the physics of interacting bosonic
systems since the realization of Bose-Einstein condensation in trapped alkali
gases\cite{leggett}. Recently, much attention has focused on bosons with 
internal degrees of freedom. These discrete internal degrees of freedom come
from the (nearly) degenerate hyperfine levels or other sources that give rise
to multiple components for the bosons. Indeed, recent 
studies\cite{ho,yin,pu,search,rojo,stoof,cooper} demonstrated
that the multi-component bosonic 
systems have much richer physics than their single-component 
counterpart due to the additional degrees of freedom.

Despite intensive research, there has been relatively few rigorous results on 
these interesting systems. In this paper we study an example of such systems,
namely a two-component bosonic system with component-independent mass and
interactions\cite{rb}.
Obviously the two components of the boson can be mapped to a pseudo-spin
degree of freedom with spin quantum number $S=1/2$ ({\em i.e.}, pseudo-spin 
``up" state representing one component, pseudo-spin ``down" state 
representing the other component), and the interaction is 
pseudo-spin independent, thus the system is invariant under pseudo-spin 
$SU(2)$ rotation. We prove rigorously that for a large class of real 
Hamiltonians, the ground state of the system is a fully polarized
pseudo-spin ferromagnet. The origin of the ferromagnetism is neither 
spin-dependent interaction (as there is none in our model), nor interparticle
interaction as Coulomb interaction driven ferromagnetism
in electronic systems; instead it is due to the fact the kinetic energy term
of a bosonic system forbids the ground state wave function to change sign, thus
forcing a totally symmetric spatial wave function and hence a totally symmetric
(and fully polarized) 
pseudospin wave function. Thus the ferromagnetism is driven by
{\em kinetic} energy in our system. We will also study the ferromagnetic spin
wave spectra of the system under various conditions.
For simplicity we will refer to the pseudospin of the bosons as spin from now
on.

Consider the following Hamiltonian describing $N$ identical bosons:
\beq
H=\sum_i\left[{{\bf p}_i^2\over 2m}+U({\bf r}_i)\right]
+\sum_{i<j}V({\bf r}_i-{\bf r}_j).
\label{contham}
\eeq
Here $U$ is the trapping potential and $V$ is the two-particle interaction.
Our proof can also be generalized to cases
with multi-body (three and above) interactions when present.
Since $H$ is 
spin independent, we have $[H, {\bf S}_i]=0$, where ${\bf S}_i$ is the spin
operator of $i$th particle. In particular, it possesses the global
$SU(2)$ symmetry:
$[H, {\bf S}_{tot}]=0$,
with ${\bf S}_{tot}=\sum_i{\bf S}_i$.
Thus the eigenkets of $H$ may be chosen to be simultaneous eigenkets of
${\bf S}_{tot}^2$ and $S^z_{tot}=(N_\uparrow-N_\downarrow)/2$.

We now proceed by considering the ground state of an {\em enlarged}
Hilbert space for
$H$, namely that of $N$ {\em distinguishable} particles, {\em i.e.}, no
permutation symmetry is required for the wave function. Since $H$ is spin 
independent and commutes with $S_i^z$ for every $1 \le i\le N$, 
in this {\em enlarged} 
Hilbert space we can always choose the eigenkets of $H$ to be 
simultaneous eigenkets of $S_i^z$, so that the wave functions take a factorized
form\cite{note}:
\beq
\psi({\bf r}_1, \sigma_1, \cdots, {\bf r}_N, \sigma_N)
=\phi({\bf r}_1, \cdots,
{\bf r}_N)\chi(\sigma_1, \cdots, \sigma_N).
\eeq
Since the eigen energy is independent of the spin wave function,
we only need to focus on the spatial wave function of the ground state: 
$\phi_0({\bf r}_1, {\bf r}_2, \cdots,
{\bf r}_N)$. Since $H$ is real, $\phi_0$ can be chosen to be real as well.
We now prove that $\phi_0$ can be chosen to be real and non-negative, using a
well-known trick. If 
$\phi_0$ contains both positive and negative parts, we can then construct a 
trial wave function that is non-negative:
$\tilde\phi_0({\bf r}_1, {\bf r}_2, \cdots,
{\bf r}_N)=|\phi_0({\bf r}_1, {\bf r}_2, \cdots,
{\bf r}_N)|$, and show that 
$\langle \tilde\phi_0|H|\tilde\phi_0\rangle$ = $\langle \phi_0|H|\phi_0\rangle$.
It is easy to see that the expectation values of the potential terms are the
same; the expectation values of the kinetic energy term are also the same 
because the integrand is the exactly the same everywhere $\phi_0\ne 0$, while
the singularity of $\nabla^2 \tilde\phi_0$ where $\phi_0=0$ is not strong enough
to overcome the speed that $\tilde\phi_0\rightarrow 0$ and makes no 
contribution. Thus we find from variational principle that  
the ground state energy $E_0 \le \langle \phi_0|H|\phi_0\rangle$. The equal 
sign holds only when $\tilde\phi_0$ happens to be the true ground state, which
is {\em not} the case under
generic situations. Thus generically the ground state wave function $\phi_0$
is non-negative definite, and moreover, positive definite for non-singular
$U$ and $V$\cite{feynman}. 
This also implies that the ground state must be 
{\em non-degenerate}, because if there existed two different ground state wave 
functions that are both non-negative, it is always possible to construct new
ground state wave functions that have both positive and negative parts by
making linear combination of the two. 

Having established that $\phi_0$ is non-degenerate and non-negative definite
under generic situations, it follows that it must also be totally symmetric:
$P_{ij}\phi_0 = \phi_0$, where $P_{ij}$ is the permutation operator for 
particles $i$ and $j$. This is because $[P_{ij}, H]=0 $ thus a 
{\em non-degenerate} eigenket of $H$ must also be an eigenket of $P_{ij}$, and
the fact $\phi_0$ is non-negative ensures the eigenvalue can only be $+1$.

We can now prove that the ground state wave function
of the spin-1/2 {\em bosonic} system must take the form:
\beq
\psi_0^B({\bf r}_1, \sigma_1, \cdots, {\bf r}_N, \sigma_N)
=\phi_0({\bf r}_1, \cdots,
{\bf r}_N)\chi_{N/2}(\sigma_1, \cdots\sigma_N),
\label{ground}
\eeq
where $\chi_{N/2}(\sigma_1, \cdots\sigma_N)$ is a spin wave function with
quantum number $S_{tot}=N/2$.
First of all, $\psi^B_0$ is symmetric under permutation, since both 
$\phi_0$ and $\chi_{N/2}$ are symmetric under permutation.
Secondly, $\psi^B_0$ is an eigen wave function of $H$ with eigen value $E_0$, 
thus it must be one of the ground states.
%Thirdly, since $\phi_0$ is the non-degenerate spatial eiegen wave function of
%$H$, it must be the common factor of all term of 
Lastly but most importantly, we need to prove the lowest energy state with 
other $S_{tot}=S < N/2$ 
quantum numbers have energies higher than $E_0$. Without losing
generality we can choose to focus on states with $S_{tot}^z=S_{tot}=S$.
Such a state takes the following form:
\beqarr
&&\psi_S^B({\bf r}_1, \sigma_1, \cdots, {\bf r}_N, \sigma_N)
=\sum_P\phi_S({\bf r}_{P(1)},\cdots, {\bf r}_{P(N)})\nonumber\\
&&\times |\uparrow_{P(1)}\cdots\uparrow_{P(N_1)};
\downarrow_{P(N_1+1)}\cdots\downarrow_{P(N)}\rangle,
\label{S}
\eeqarr
where $P$ stands for permutation, and $N_1=N/2+S$ is the number of up
spin particles. In order for $\psi_S^B$ to be an eigen wave function of $H$ with
eigen value $E$, $\phi_S$ must be an eigen spatial wave function of $H$ with
eigen value $E$, which can be chosen to be real. 
Now we prove $E > E_0$. Since $\psi_S^B$ is a common eigen ket
of ${\bf S}_{tot}^2$ and $S^z_{tot}$ with $S^z_{tot}=S_{tot}$, we have
\beq
S_{tot}^+|\psi_S^B\rangle=\sum_iS_i^+|\psi_S^B\rangle=0.
\label{ann}
\eeq
Since $S_{tot}^+$ does {\em not} annihilate any term of 
$|\psi_S^B\rangle$ expanded as in eq. (\ref{S}), and matrix elements of $S^+_i$
are real and positive in the $S^z$ representation, in order for 
eq. (\ref{ann}) to be valid $\phi_S$ must contain both positive and negative 
parts. Thus $E > E_0$.
We have thus proved that the ground state of the bosonic system described by the
Hamiltonian (\ref{contham}) must have
$S_{tot}=N/2$ with degeneracy $2S_{tot}+1=N+1$; 
if there exists an {\em infinitesimal} magnetic field, the degeneracy will be
lifted and the spins will be fully-polarized along the direction of the field. 

Using similar methods
one can also prove the same is true for lattice Hamiltonians of the form:
\beq
H=-\sum_{(ij),\sigma}t_{ij}(b_{i\sigma}^\dagger b_{j\sigma}
+b_{j\sigma}^\dagger b_{i\sigma})+\sum_iU_in_i+\sum_{(ij)}V_{ij}n_in_j,
\label{latham}
\eeq
where $b_{i\sigma}$ is the boson annihilation operator for site $i$ and spin
component $\sigma$, $t_{ij}$ are real and {\em positive}
hopping matrix elements,
$V_{ij}$ is the two-body interaction potential, and 
$n_i=b_{i\uparrow}^\dagger b_{i\uparrow}+b_{i\downarrow}^\dagger b_{i\downarrow}$
is the total boson number on site $i$. The lattice structure is not important
and hence not specified.

We emphasize our proof does {\em not} assume the presence of BEC; in fact it is
valid in the absence of BEC as well, and we will discuss an example of this
situation later. On the other hand it is crucial that the Hamiltonian $H$ is
{\em real}, so that the eigen wave functions can be chosen to be real; the
proof is no longer valid when the Hamiltonian $H$ is complex, for example 
when the bosons are charged and moving in a magnetic field (with {\em no}
Zeeman coupling so that the SU(2) symmetry is present), or the bosons are 
placed in a rotating trap. 

A physical consequence of the ferromagnetism of the ground state is the presence
of ferromagnetic spin wave excitations. In the following we study the spin wave
spectra of two different situations, one in the continuum and the other in a 
lattice model.

Consider a continuum system described by the Hamiltonian Eq. (\ref{contham})
with $U=0$, so that the system is translationally invariant and all eigenkets
of $H$ can be chosen to have a momentum quantum number ${\bf k}$. A natural
variational wave function for a ferromagnetic spin-wave state takes the form
\beq
|{\bf k}\rangle=S_{\bf k}^-|0\rangle,
\eeq
where $|0\rangle$ is the ground state with $S_{tot}=S_{tot}^z=N/2$, and
\beq
S_{\bf k}^-=(1/\sqrt{N})\sum_je^{i{\bf k}\cdot{\bf r}_j}S_j^-.
\eeq
This is, of course, a straightforward generalization of the 
single-mode-approximation (SMA) for {\em spinless} bosons\cite{feynman2}.
Thus a variational approximation (or upper bound) for the spin-wave spectrum
is 
\beq
E(k)=f(k)/s(k)={\hbar^2k^2\over 2m},
\eeq
where 
\beq
f(k)=\langle {\bf k}|(H-E_0)|{\bf k}\rangle ={\hbar^2k^2\over 2m}
\eeq
is the same oscillator strength as in SMA,
while here
\beq
s(k)=\langle {\bf k}|{\bf k}\rangle 
= \langle 0|S_{\bf -k}^+ S_{\bf k}^-|0\rangle=1
\eeq
is the {\em spin} structure factor. In the SMA for spinless bosons the 
collective mode spectrum is linear for small $k$ because the {\em density}
structure factor is linear in $k$ for small $k$\cite{feynman2}; 
here the spin-wave spectrum 
is quadratic as expected for Heisenberg ferromagnets, and we find that it is
bounded from above by the single particle spectrum.

For bosons moving in a periodic potential\cite{greiner},
the system can be appropriately described by a tight-binding lattice Hamiltonian
of the form (\ref{latham}). In particular, if the periodic potential is 
strong enough the system loses BEC and becomes a Mott insulator\cite{greiner}. 
In this case
the charge degree of freedom of the bosons are frozen, but the spin degree of 
freedom are still active and as we show below, they can be described by the 
Heisenberg model with {\em ferromagnetic} couplings. For simplicity we consider
a boson Hubbard model with nearest neighbor hopping and onsite repulsion,
on a simple cubic lattice:
\beqarr
H&=&-t\sum_{<ij>,\sigma}(b_{i\sigma}^\dagger b_{j\sigma}
+b_{j\sigma}^\dagger b_{i\sigma})+\sum_{i}(U/2)(n_i-1)^2\nonumber\\
&=&-t\sum_{<ij>,\sigma}(b_{i\sigma}^\dagger b_{j\sigma}
+b_{j\sigma}^\dagger b_{i\sigma})-(U/2)\sum_{i} n_i \nonumber\\
&+&\sum_{i}Un_i(n_i-1)/2+const.,
\eeqarr
and assume the number of bosons is the same as the number of lattice sites, 
thus there is one boson on every site in average. For very large $U/t$ the 
bosons can not hop from one site to another due to the large potential energy
cost, and the system is in the Mott insulator phase with {\em no} BEC.
A straightforward second-order perturbation calculation in $t/U$ yields an
effective Heisenberg spin Hamiltonian for the system:
\beq
H=J\sum_{<ij>}({\bf S}_i\cdot {\bf S}_j+3/4)
\eeq
with {\em ferromagnetic} coupling:
\beq
J=-4t^2/U.
\eeq
And the spin-wave spectrum 
takes the {\em exact} form 
\beq
E({\bf k})=(8t^2/U)(3-\cos k_x-\cos k_y-\cos k_z)
\eeq
in this limit. This is exactly the opposite for the Hubbard model at 
half-filling for electrons, where the large $U/t$ limit leads to a Heisenberg
{\em antiferromagnet}. We note in passing that in {\em electronic} systems, 
ferromagnetism is usually associated with {\em itinerant} electrons while
localized electrons usually give rise to antiferromagnetism; thus two-component
bosons trapped to lattice potentials provide new opportunities to study lattice
ferromagnets.

We now turn our discussion to the relation between our work and existing 
theoretical work on this and related fields. 

Ho and Yin\cite{yin} considered the general spin structure of Bose-Einstein
condensates of bosons with arbitrary spin. In their work they assumed weakly 
interacting bosons thus to a very good approximation, all bosons occupy the
lowest-energy orbital state. They pointed out for spin-1/2 bosons due to 
Bose statistics, the bosons must also occupy the same spin state and the 
system is a ferromagnet, and very appropriately termed the ferromagnetic
state ``statistical ferromagnet". Our work has considerable overlap with theirs
in spirit, and certainly comes to the same 
conclusion as theirs. On the other hand our proof is rigorous and more general;
it applies to interacting bosons with arbitrary interaction strength and in
particular, strongly interacting bosonic systems without BEC. We have thus     
generalized the statistical ferromagnet to a much wider class of bosonic
systems. 

Rojo\cite{rojo} studied two-component bosonic systems with 
{\em component-dependent}
interactions, in which the intercomponent interaction is always more repulsive
or less attractive than the intracomponent interaction. He showed that in this
case the ground state is ``fully polarized" in the sense that only bosons of
one component appear. In the spin analogy the interaction he considered has
Ising anisotropy, and the system is an Ising ferromagnet. Thus his work is 
complementary to ours, which studies a Heisenberg ferromagnet. By the same
token, in his model if the intercomponent interaction were
less repulsive or more
attractive than the intracomponent interaction, the interaction would have 
XY anisotropy and the system would be an XY ferromagnet. He further found 
that as the two interactions become closer and closer to each other (or one
is approaching the Heisenberg point), the velocity of one of the two linear
collective modes vanishes. This, as we see here, is because at the 
Heisenberg point the spin wave spectrum has {\em quadratic} dispersion at long
wave length.

Li and coworkers\cite{li} have studied one-dimensional SU(2) bosons with 
$\delta$-function interactions. They used Bethe ansatz to obtain the ground
and low-lying state spectra and quantum numbers, and showed the ground state
is fully spin polarized and there is a quadratic spin wave mode. Their results
certainly agree with the general conclusions of the present work.

Pu {\em et al.}\cite{pu} studied possibility of ferromagnetic states in bosonic
systems originated from dipole-dipole interactions. In their model the 
ferromagnetism comes from {\em interactions} 
that are magnetic in the first place, which is very different from our case,
where the ferromagnetism is essentially {\em kinetic energy} driven.

Thus far, our discussions have been focusing on bosonic systems. Fermionic
systems are of course very different, due to the difference in statistics. 
On the other hand in two-dimensions (2D) the statistics of the particles can be 
transformed through flux attachment transformations\cite{zhang}. 
Thus our results also have
implications on electrons in a strong magnetic
field in the quantum Hall regime.
For example, it gives a new perspective to the spin/pseudospin ferromagnetism
for single/bi-layer quantum Hall systems at Landau level filling factor at 
$\nu=1$\cite{sondhi,yang}, 
as the electrons can be mapped to bosons carrying one flux quantum,
and the flux carried by the bosons {\em cancel} that of the external magnetic
field {\em in average}, thus at the mean-field level these systems are mapped
onto boson systems at {\em zero} magnetic field, precisely the system we study
here! Using the same transformation one can also map spin-1/2
bosons in a magnetic
field with $\nu=1$ to fermions with zero field, 
where one expects a {\em singlet} ground state. This has indeed been found to be
the case in finite-size studies, for Coulomb interaction\cite{rezayi}.

We thank T.-L. (Jason) Ho for useful discussions and comments on the manuscript.
This work was supported by NSF Grant No. DMR-9971541, and the A. P. Sloan
Foundation (KY), and trans-century project of China
Ministry of Education and NSFC No.1-9975040 (YQL).


\begin{references}
\bibitem{note} This is only true for the enlarged Hilbert space, but not true
for the bosonic or fermionic system where specific permutation symmetry is
required for the wave functions.
\end{references}
\end{document}